\documentclass[%
 reprint,
 amsmath,amssymb,
 aps,
prl,
floatfix,
longbibliography,
]{revtex4-2}

\usepackage{graphicx}
\usepackage{subcaption}
\usepackage{dcolumn}
\usepackage{bm}
\usepackage[english]{babel}
\usepackage{upgreek}
\usepackage{ifthen}

\usepackage{comment}
\usepackage{tikz}
\usepackage{pgfplots}
\pgfplotsset{compat=1.17}

\newcommand{\bs}[1]{\boldsymbol{#1}}

\newcommand{\mc}[1]{\mathcal{#1}}

\newcommand{\half}[0]{\frac{1}{2}}

\newcommand{\paren}[1]{\left(#1\right)}

\newcommand{\avg}[1]{\langle #1 \rangle}

\newcommand{\kev}{\text{keV}}

\newcommand{\mr}[1]{\mathrm{#1}}

\newcommand{\vth}{v_\mr{th}}

\newcommand{\expp}[1]{\exp\left\{#1\right\}}

\newcommand{\kn}[0]{\mathrm{Kn}}
\newcommand{\gk}[0]{\mathrm{Gk}}

\newcommand{\nuth}{\nu_0}

\newboolean{showHeadings}
\setboolean{showHeadings}{false} 

\newboolean{showEdits}
\setboolean{showEdits}{false}

\newcommand{\deletedtext}[1]{\ifthenelse{\boolean{showEdits}}{\textcolor{red}{#1}}{}}
\newcommand{\addedtext}[1]{\ifthenelse{\boolean{showEdits}}{\textcolor{blue}{#1}}{#1}}
\newcommand{\movedtext}[1]{\ifthenelse{\boolean{showEdits}}{\textcolor{gray}{#1}}{#1}}

\newenvironment{addedblock}
  {\ifthenelse{\boolean{showEdits}}{\color{blue}}{}}
  {\ifthenelse{\boolean{showEdits}}{}{}}

\newenvironment{movedblock}
  {\ifthenelse{\boolean{showEdits}}{\color{gray}}{}}
  {\ifthenelse{\boolean{showEdits}}{}{}}

\begin{document}

\preprint{APS/123-QED}

\title{Enhancement to Fusion Reactivity in Sheared Flows}

\author{Henry Fetsch}
 \email{hfetsch@princeton.edu}
\author{Nathaniel J. Fisch}%
\affiliation{Department of Astrophysical Sciences, Princeton University, Princeton, NJ}

\date{\today}

\begin{abstract}
    Sheared flow increases the reactivity of fusion plasma. In unmagnetized plasma with flow gradients comparable to the mean free path of reacting ions, fusion reactivity can be more than doubled. 
    The effect is of particular relevance to inertial confinement fusion (ICF), where it allows implosion kinetic energy to contribute to the fusion burn even before thermalizing. 
    In fast ignition, this allows remarkable energy savings because colder fuel stops alpha particles more quickly, enabling ignition in a smaller hot spot. 
\end{abstract}

\maketitle


\ifthenelse{\boolean{showEdits}}{
    \noindent Redlined manuscript:
    \\ \textcolor{red}{Deleted text in red.}
    \\ \textcolor{blue}{Added text in blue.}
    \\ \textcolor{gray}{Moved text in gray.}
}{}

\ifthenelse{\boolean{showHeadings}}{
    \begin{center}
        \textit{Introduction}
    \end{center}

}{\textit{Introduction -- }}
An axiom or belief in inertial confinement fusion (ICF) is that imploding plasma must thermalize to achieve high fusion reactivity \cite{Lindl_1995, Rosen_2024}. 
The alternative would have the implosion energy directed into rapidly swirling turbulent eddies of lower temperature.  But, since fusion depends only on the relative motion of colliding ions, any turbulent energy is useless -- in the local fluid frame of reference, only temperature produces relative motion of the reactants \cite{Clark_et_2019,Davidovits_Fisch_2019_2dcomp}.  Hence, thermalization is axiomatically crucial.  This work overturns that belief.  Some energy in turbulent eddies, rather than temperature, can in fact produce higher fusion reactivity.

This \textit{shear flow  reactivity enhancement effect} occurs because, in fact, relatively few suprathermal ions, with mean free paths many times longer than those of thermal ions, govern fusion reactivity. Even when a hydrodynamic description is appropriate for thermal particles, the tail of the distribution often demands a kinetic treatment. In such cases, fusion reactivity $\avg{\sigma v}$ becomes a nonlocal functional of fluid properties within a few fast-ion mean free paths \cite{Molvig_Hoffman_Albright_Nelson_Webster_2012,Albright_Molvig_Huang_Simakov_Dodd_Hoffman_Kagan_Schmit_2013,Rinderknecht_Amendt_Wilks_Collins_2018}. 

Surprisingly, the effect of inhomogeneous flow on reactivity has so far not been considered. 
In this Letter, we show for the first time that sheared flow directly enhances fusion reactivity\addedtext{ and that the effect can be huge}.
Inhomogeneous flow can, of course, modify reactivity by compressing and heating patches of fluid, but, remarkably, even solenoidal flow enhances reactivity through the nonlocal effect of fast ions.\deletedtext{ICF produces the exotic conditions capable, in extreme cases, of multiplying reactivity severalfold.}\deletedtext{In systems near ignition, even modest changes in reactivity can affect yield dramatically.} Furthermore, exchanging thermal for turbulent energy allows fuel to ignite at lower temperature; leveraging this fact in fast-ignition designs allows ignitor energy to be reduced by more than half. 
Sheared flow generates an even larger enhancement for ``advanced fuels,'' such as D\textsuperscript{3}He and p\textsuperscript{11}B, potentially enabling ICF using these \deletedtext{fuels}\addedtext{aneutronic reactions}. 



\begin{figure}
    \centering
    \begin{tikzpicture}
        \begin{axis}[
            width = 0.8\columnwidth,
            xlabel={$v_x$},
            ylabel={$f(z,v_x)$},
            ymin=0, ymax=4,
            xmin=-6, xmax=6,
            axis lines=left,
            xtick = \empty,
            ytick = \empty
        ]
            \addplot[
                domain=-5:5, 
                samples=100, 
                smooth, 
                very thick,
                color=blue
            ] {exp(-((x-1)^2)/2) + 0.25}; 
            \node[black, anchor=east] at (-5, 0.25) {$z_2$};
            
            \addplot[
                domain=-5:5, 
                samples=100, 
                smooth, 
                very thick,
                color=blue
            ] {exp(-((x+0)^2)/2) + 1.5}; 
            \node[black, anchor=east] at (-5, 1.5) {$z_1$};
            
            \addplot[
                domain=-5:5, 
                samples=100, 
                smooth, 
                very thick,
                color=blue
            ] {exp(-((x+1)^2)/2) + 2.75}; 
            \node[black, anchor=east] at (-5, 2.75) {$z_0$};

            \addplot[
                domain=-5:5, 
                samples=100, 
                dotted, 
                very thick,
                color=purple
            ] {1*exp(-((x+1)^2)/2) + 2.75 + 2*exp(-2/(0.1+abs(x+0.5)))*(exp(-((x+0)^2)/2)-exp(-((x+1)^2)/2))}; 

            \draw[->, dashed, thick, red] (axis cs:1, {exp(-((1+0)^2)/2) + 1.5}) -- (axis cs:1, {exp(-((1+1)^2)/2)+0.25 + 2.75});
            \draw[->, dashed, thick, red] (axis cs:2, {exp(-((2-1)^2)/2) + 0.25}) -- (axis cs:2, {exp(-((2+1)^2)/2)+0.1 + 2.75});

            \draw[->, very thick, blue] (axis cs:-5, 0.5) -- (axis cs:-2.5, 0.5);
            \draw[->, very thick, blue] (axis cs:-5, 1.75) -- (axis cs:-3.5, 1.75);
            \draw[->, very thick, blue] (axis cs:-5, 3) -- (axis cs:-4.5, 3);

        \end{axis}
    \end{tikzpicture}
    \caption{Fast particles crossing a flow gradient perturb the distribution, particularly on the tail.}
    \label{fig_sketch_sheared_flow}
\end{figure}
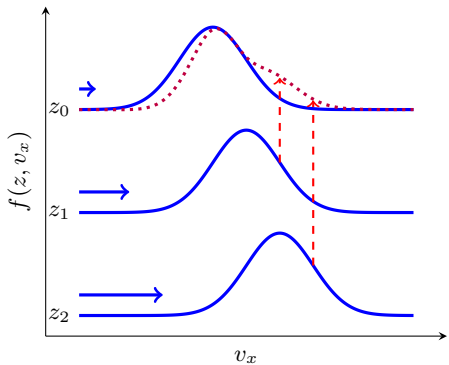

\ifthenelse{\boolean{showHeadings}}{
    \begin{center}
        \textit{Physical picture}
    \end{center}

}{\textit{Physical picture -- }}
The effect can be understood simply in physical terms as follows. 
Consider, as sketched in Fig.~\ref{fig_sketch_sheared_flow}, a fluid flowing in the $x$ direction and sheared in the $z$ direction. The distribution function at each point is approximately a drifting Maxwellian. 
A particle sampled from the thermal bulk at $z_1$ ends up closer to the tail if it reaches $z_0$ without colliding. For a thermal particle sampled at $z_2$, the effect is more pronounced. 
Some particles will have smaller velocities in their new frame, but the overall effect is to broaden the tail. 
Neglecting slowing by electrons, the mean free path $\lambda$ for fast ions scales as $\lambda \propto w^{4}$, where $w$ is the velocity in the local fluid frame, so faster ions communicate further across the flow gradient. 
Thermal ions travel a relatively short distance $\lambda_\mr{th}$ between collisions; the dynamics of these ions give rise to viscosity. 
Fast particles, however, travel beyond viscous length scales, potentially crossing large flow differentials.


For a gradient length scale $L$, a useful figure is the Knudsen number $\kn \doteq \lambda_\mr{th}/L$. 
For Maxwellian ions, the reaction rate is peaked at ${v_* \propto \vth(E_G/T)^{1/6}}$, where the Gamow energy $E_G$ gives the energy scale of the Coulomb barrier, $T$ is the temperature (in units of energy), and $\vth$ is the thermal velocity. In many cases, ${v_* \gg \vth}$. Because most reacting particles have velocities near $v_*$, the ``Gamow mean free path'' $\lambda_*$ is much longer than $\lambda_\mr{th}$.
The `Gamow-Knudsen' number ${\gk \doteq \lambda_*/L = (v_*/\vth)^4 \kn}$ of McDevitt \textit{et al.} 
thus better captures the effect of gradients on reactivity \cite{McDevitt_Tang_Guo_2017, Molvig_Hoffman_Albright_Nelson_Webster_2012}.

When ions cross a flow gradient, they attain some drift velocity in their new frame; we show here that even a small drift has a large effect on reactivity. 
Consider a one-dimensional Maxwellian distribution of ions with thermal velocity $\vth$ reacting with some stationary background species. Suppose that the fusion power $P$ is proportional to the number of particles at $v = v_*$ (this is a reasonable estimate for strongly resonant reactions, while for others it is merely illustrative), so in equilibrium ${P^\mr{(eq)}\propto \expp{-\half v_*^2/\vth^2}}$. 
Suppose that we impart to the ions a drift velocity ${u \ll \vth\sqrt{\vth/v_*}}$. The fusion power is now approximately ${P^\mr{(drift)} \propto \exp(-\half v_*^2/\vth^2 + v_* u/\vth^2)}$. The drift energy could instead have been used for heating. If the drifting ions are one quarter of total particles, heating produces a new thermal velocity $\vth'$ such that ${{\vth'}^2 = \vth^2 + \frac{1}{4}u^2}$, yielding fusion power ${P^\mr{(heat)} \propto \exp(-\half v_*^2/\vth^2 + \frac{1}{8} v_*^2 u^2 / \vth^4)}$. Denoting by $\Phi$ the enhancement factor relative to $P^\mr{(eq)}$, we have
\begin{align}
    \Phi^{\mr{(drift)}} \sim e^{\gamma} && \mr{and} && \Phi^{\mr{(heat)}} \sim e^{\gamma^2/8} ,
\end{align}
where $\gamma \doteq v_* u/\vth^2$ is the single parameter governing the size of the effect. 
For $\gamma \gtrsim 1$, fusion power is multiplied manifold even though $u$ is small. For $\gamma < 8$, energy is better spent on drift than on heating. 
By this principle, beam-target fusion offers a high reaction rate \cite{Mikkelsen_1989,Kirov_et_2021}, but with low efficiency since most beam ions scatter before fusing. Anisotropic or counterstreaming Maxwellians likewise increase reaction rates \cite{Dawson_Furth_Tenney_1971,Kolmes_Mlodik_Fisch_2021,Nath_Majumdar_Kalra_2013,Xie_Tan_Luo_Li_Liu_2023}, but energy must be expended to keep such distributions out of equilibrium, which prohibits use in a fusion reactor \cite{Rider_1997}. 
The shear flow reactivity enhancement is fundamentally different because populations are separated in space. 
The increase in mean free path with velocity leads to a drifting population of fast ions while the colder thermal bulk is more collisional and therefore relatively inviscid, leading to slow dissipation. \addedtext{This effect is peculiar to plasma, rather than other fluids, because in plasma faster particles have longer mean free paths.}

\ifthenelse{\boolean{showHeadings}}{
    \begin{center}
        \textit{Kinetic model}
    \end{center}

}{\textit{Kinetic model -- }}
To isolate the effect of flow shear, we consider uniform, planar $x$-directed flows separated by a transition layer of width $L$ at $z=0$. Away from this layer, the flow field $\bs u$ is given by
\begin{equation}
\label{eq_flow_profile}
    \bs u(z) = 
    \begin{cases}
        u_0 \bs {\hat x} & z \ll L 
        \\
        0 & z \gg L 
    \end{cases} 
\end{equation}
where $u_0$ is a constant. 
We first consider regimes in which
\begin{equation}
    \label{eq_kn_ordering}
    \kn \ll 1 \ll \gk ,
\end{equation}
permitting a hydrodynamic description: thermal particles, which are responsible for the majority of viscous dissipation, have short mean free paths. On the other hand, particles near the Gamow peak travel far beyond the layer before scattering. If we are exclusively interested in the transport of energetic particles, we can treat the layer as a point discontinuity. In the numerical work below, we will relax this assumption to consider systems in which the scale separation \eqref{eq_kn_ordering} is not so large.

For simplicity, we consider a single ion species with mass $m$, uniform density $n$, and uniform temperature $T$. Electrons form a neutralizing background. The distribution function $f$ evolves according to
\begin{equation}
\label{eq_kinetic_eq}
    \frac{\partial f}{\partial t}  + \bs v \cdot \nabla f + \bs a \cdot \frac{\partial f}{\partial \bs v}  = \mc C[f, f] ,
\end{equation}
where $\bs v$ is the velocity in the laboratory frame, $\bs a$ is the net acceleration, and $\mc C$ is a bilinear collision operator.

Let $\bs w \doteq \bs v - \bs u$ be the peculiar velocity and ${\vth \doteq (T/m)^{1/2}}$ be the thermal velocity. By \eqref{eq_kn_ordering}, the bulk of the distribution is approximately a Maxwellian $f_M$, while the tail may deviate significantly. We approximate collisions as occurring between test particles and a thermal background, adopting a Bhatnagar-Gross-Krook (BGK) operator, \textit{viz.} ${\mc C = - \nu_\mr{ii} (f - f_M)}$ 
where $\nu_\mr{ii} = \nuth$ for $w \leq \vth$, $\nu_\mr{ii} = \nuth(\vth/w)^3$ for $w > \vth$, and $\nuth$ is a constant. We finally assume steady state (${\partial_t f = 0}$) and that all spatial variation is in the $z$ direction. 
Treating the shear layer as a step function, the solution is
\begin{equation}
    \label{eq_f_approx}
    f(z > 0) \sim \begin{cases} 
        \begin{aligned}
            & \frac{\hat f}{(2\pi)^{3/2}\vth^3} \Bigg[ e^{-\frac{z\nu_\mr{ii}}{w}}\expp{-\frac{|\bs w - u_0 \bs{\hat x}|^2}{2\vth^2}  } \\
            & +  \paren{1-e^{-\frac{z\nu_\mr{ii}}{w}}}\expp{-\frac{|\bs w |^2}{2\vth^2} } \Bigg]
        \end{aligned}
          &  w_z > 0 , \\
          \begin{aligned}
            \frac{1}{(2\pi)^{3/2}\vth^{3}} \expp{-\frac{|\bs w |^2}{2\vth^2} }
          \end{aligned}
        &  w_z \leq 0 ,
        \end{cases} 
\end{equation}
where $\hat f$ is a normalization factor to conserve density. The distribution for $z < 0$ is obtained by reversing signs.

\ifthenelse{\boolean{showHeadings}}{
    \begin{center}
        \textit{Fusion reactivity}
    \end{center}

}{\textit{Fusion reactivity -- }}
For a single species, reactivity is
\begin{equation}
    \label{eq_sigma_v_general}
    \begin{split}
    \avg{\sigma v} = \half \vth \iint d^3w d^3w' \sqrt{2}p \sigma(p) f(\bs w) f(\bs w'),
    \end{split}
\end{equation}
where $\sigma$ is the fusion cross section and ${p \doteq (\bs w - \bs w')/\sqrt{2}\vth}$ is the normalized relative velocity. Assuming a constant S-factor and no resonances, $\sigma$ can be approximated by
\begin{equation}
    \label{eq_cross_section_analytic}
    \sigma(p) = \frac{A}{\vth^2 p^2} e^{-b/p} ,
\end{equation}
where $A$ and $b$ are constants (note $b \propto \vth^{-1}$). Physically, $b$ represents the height of the Coulomb barrier relative to the thermal energy ($b^2 \propto E_G/T$). In general, fusion reactions are classically forbidden and require quantum tunneling; in other words, $b$ is large. 
For DD reactions at 3~keV, for example, $b \approx 26$ \cite{Bosch_Hale_1992}. When ${b \gg 1}$, the integrand is sharply peaked, allowing \eqref{eq_sigma_v_general} to be evaluated by steepest descent. 
To leading order in $b$, we define the enhancement factor $\Phi$ as
\begin{equation}
    \label{eq_reactivity_maxwellian_phi}
    \avg{\sigma v} \sim \paren{ \sqrt{\frac{2}{3}}\frac{A b^{1/3}}{\vth} e^{-\frac{3}{2}b^{2/3}} } \Phi 
\end{equation}
so that $\Phi = 1$ for Maxwellian distributions. Starting from \eqref{eq_f_approx} and making the approximation that every excess (nonthermal) fusion reaction occurs between a particle from the perturbed distribution, i.e. the first line of \eqref{eq_f_approx}, and a particle from a Maxwellian, we have 
\begin{equation}
    \label{eq_reactivity_integral_6d}
    \begin{split}
    \Phi \sim& \half + \frac{\iint \frac{d^3p d^3s}{(2\pi)^3} \frac{e^{-b/p}}{p} e^{ - \frac{1}{2}|\bs s - \half \hat u\bs{\hat x}|^2 }  e^{-\frac{1}{2}p^2 + \half \hat u p_x - \frac{1}{8} \hat u^2}}{2\iint \frac{d^3p d^3s}{(2\pi)^3} \frac{e^{-b/p}}{p} e^{ - \frac{1}{2}s^2 }  e^{-\frac{1}{2}p^2 }} ,
    \end{split}
\end{equation}
where $s\doteq (\bs w + \bs w')/\sqrt{2}\vth$ and $\hat u \doteq \sqrt{2}u_0/\vth$.  Assuming $\hat u \lesssim 1$, evaluating \eqref{eq_reactivity_integral_6d} to leading order in $b$ yields
\begin{equation}
    \label{eq_reactivity_approx}
    \begin{split}
    \Phi \sim \half + \frac{e^{-\frac{1}{12}\hat u^2 }\sinh\paren{\half \hat u b^{1/3}}}{\hat u b^{1/3}}
    \end{split}
\end{equation}
By \eqref{eq_reactivity_maxwellian_phi}, $\avg{\sigma v}$ is a strongly decreasing function of $b$ (so an increasing function of temperature) in reactor-relevant regimes. However, by \eqref{eq_reactivity_approx}, $\Phi$ is an increasing function of $b$. Therefore, while fusion reactivity is largest at high temperature, the fractional enhancement is largest at low temperature. Additionally, the flow speed $u_0$ corresponding to a fixed $\hat u$ is smaller at low temperature. For these reasons, a promising application of strongly sheared flow is to ``jump-start'' a fusion burn in warm fuel. 
Because $b$ increases with the charge of the reacting species, the enhancement is larger for ``advanced fuels'' such as D\textsuperscript{3}He and p\textsuperscript{11}B. In plasma containing high-Z elements, whether from non-hydrogen fuels, impurities, or alpha particles, the relative value of shear flow to compared to heating is greater because additional electrons increase heat capacity but do not significantly increase flow kinetic energy.

\ifthenelse{\boolean{showHeadings}}{
    \begin{center}
        \textit{Numerical results}
    \end{center}

}{\textit{Numerical results -- }}
To improve on the analytical estimate in \eqref{eq_reactivity_approx}, we solved \eqref{eq_kinetic_eq} numerically in steady state using a BGK operator with collision frequency $\nu = \nu_\mr{ii} + \nu_\mr{ie}$. The ion-ion collision frequency $\nu_\mr{ii}$ was given by the same model as above and the ion-electron collision frequency was a constant $\nu_\mr{ie} = \nuth\sqrt{m_e/\overline{m}}$ where $m_e$, $m_D$, and $m_T$ are the electron, deuterium, and tritium masses respectively and ${\overline{m} \doteq (m_D+m_T)/2}$. An equimolar DT plasma was assumed and velocities normalized to ${\bar v_{th} \doteq \sqrt{T/\overline{m}}}$. The background was a Maxwellian with uniform temperature and with stationary flow profile
\begin{eqnarray}
    \label{eq_flow_profile_numerical}
    \bs u(z) =  \frac{u_0}{2} \paren{1 - \tanh\paren{\frac{z}{L}}} \bs{\hat x}
\end{eqnarray}
for constant flow velocity $u_0$ and gradient length scale $L$. 

\begin{figure} 
    \centering 
    \includegraphics[width=\columnwidth]{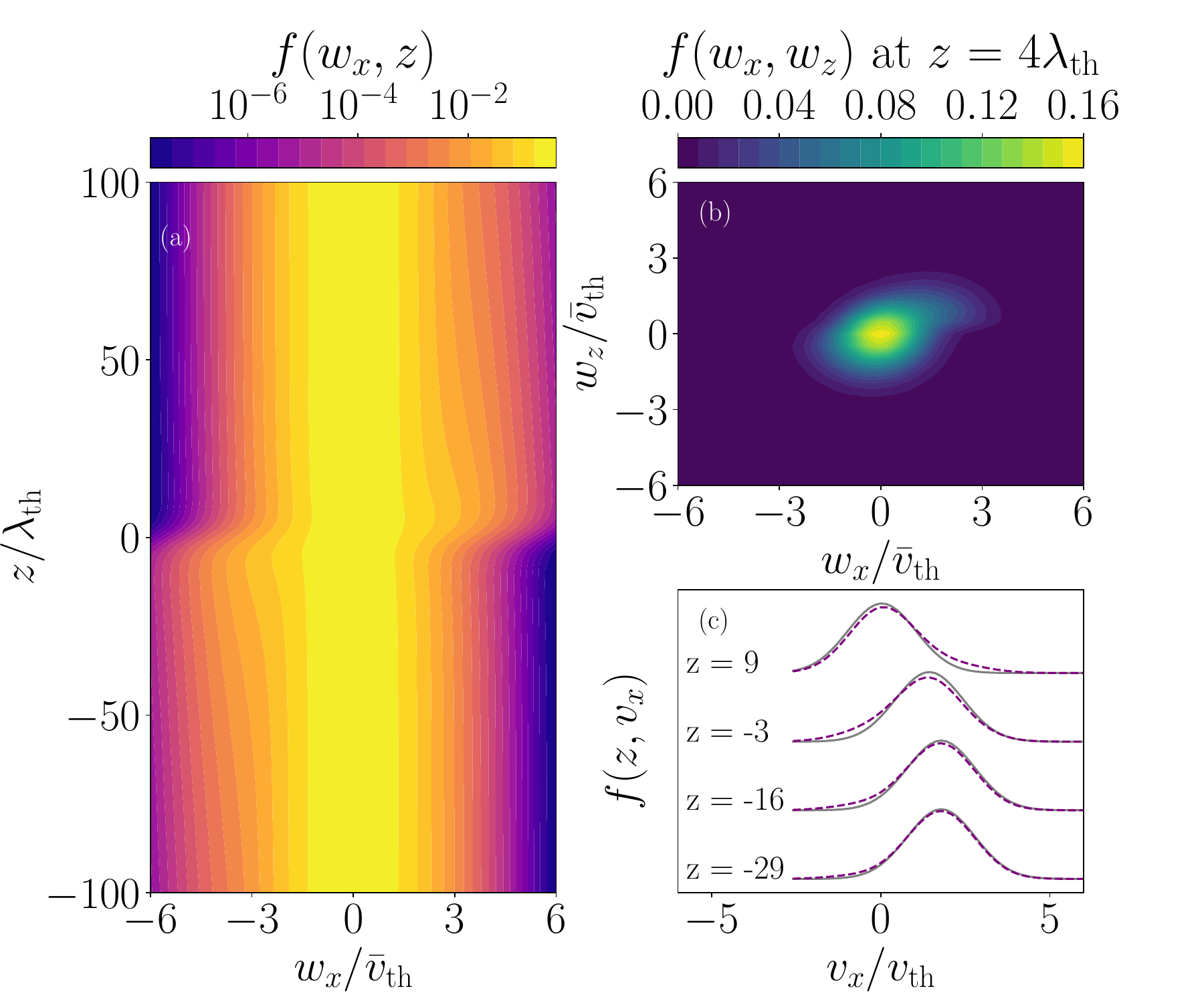}
    \caption{Distribution function for a $u_0 = 2\vth, L = 5\lambda_\mr{th}$ layer. (a) $f(z,w_x)$ showing a jump in peculiar velocity near $z=0$. (b) $f(w_x, w_z)$ showing the anisotropic distribution of fast particles. (c) $f(z, v_x)$ (purple, dotted) compared to $f_M(z, v_x)$ (grey); cf. Fig.~\ref{fig_sketch_sheared_flow}.}
    \label{fig_f_panels}
\end{figure}

Viewed in the $z-w_x$ plane (Fig.~\ref{fig_f_panels}a), the perturbation to $f$ appears as ``wings'' near $z=0$, where particles arriving from the other side of the shear layer have not yet caught up to the local flow velocity. This phenomenon affects all particles, but it persists over a much longer distance for fast particles. In the $w_x-w_z$ plane (Fig.~\ref{fig_f_panels}b), the perturbation appears as a lobe of higher phase-space density at $w_x,w_z>0$; this fast-ion population is responsible for the enhancement in reactivity.

\begin{figure}
    \centering
    \begin{subfigure}[b]{0.48\textwidth}
         \includegraphics*[width=\columnwidth]{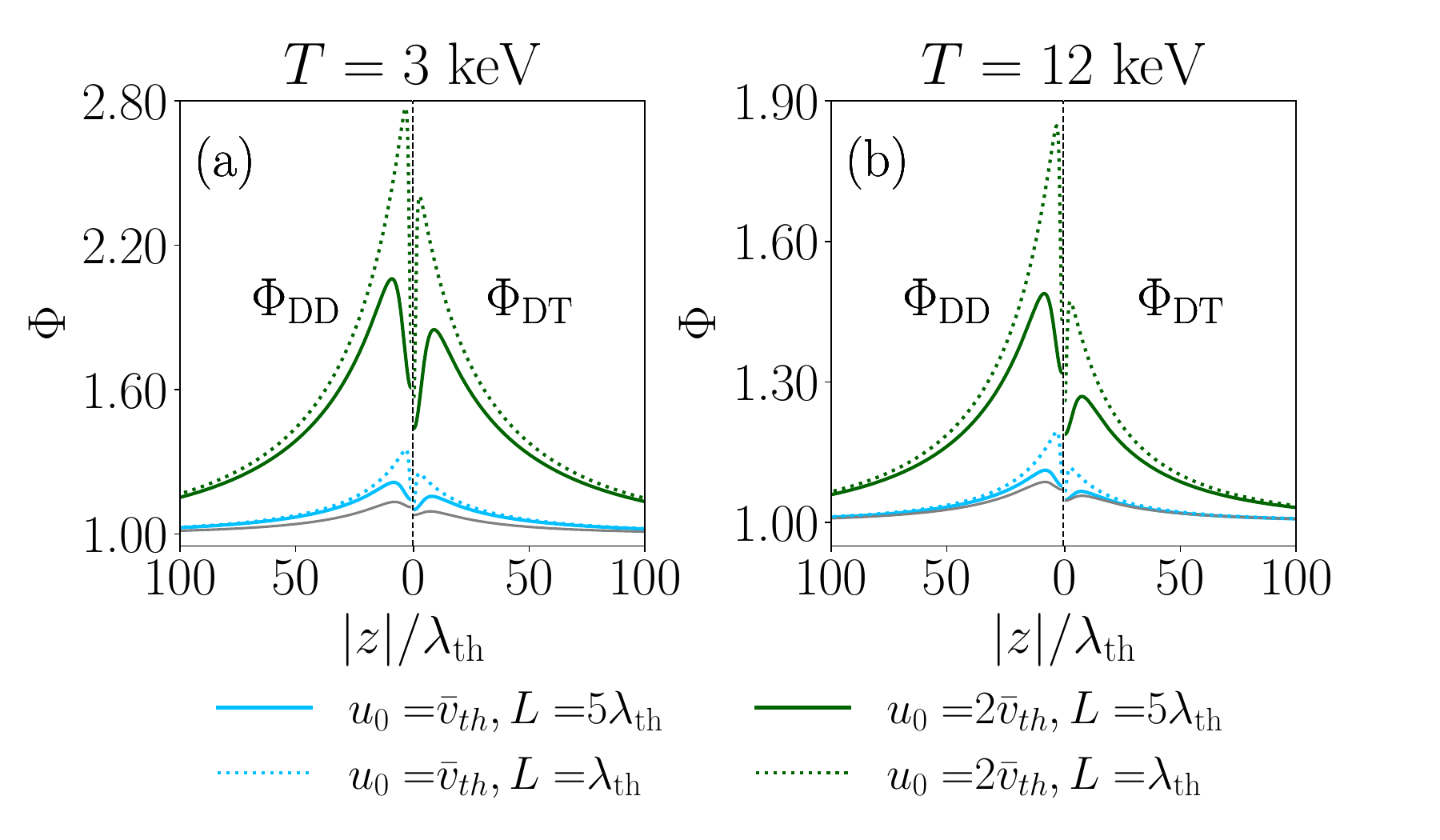}
    \end{subfigure}
    \begin{subfigure}[b]{0.48\textwidth}
     \includegraphics[width=\columnwidth]{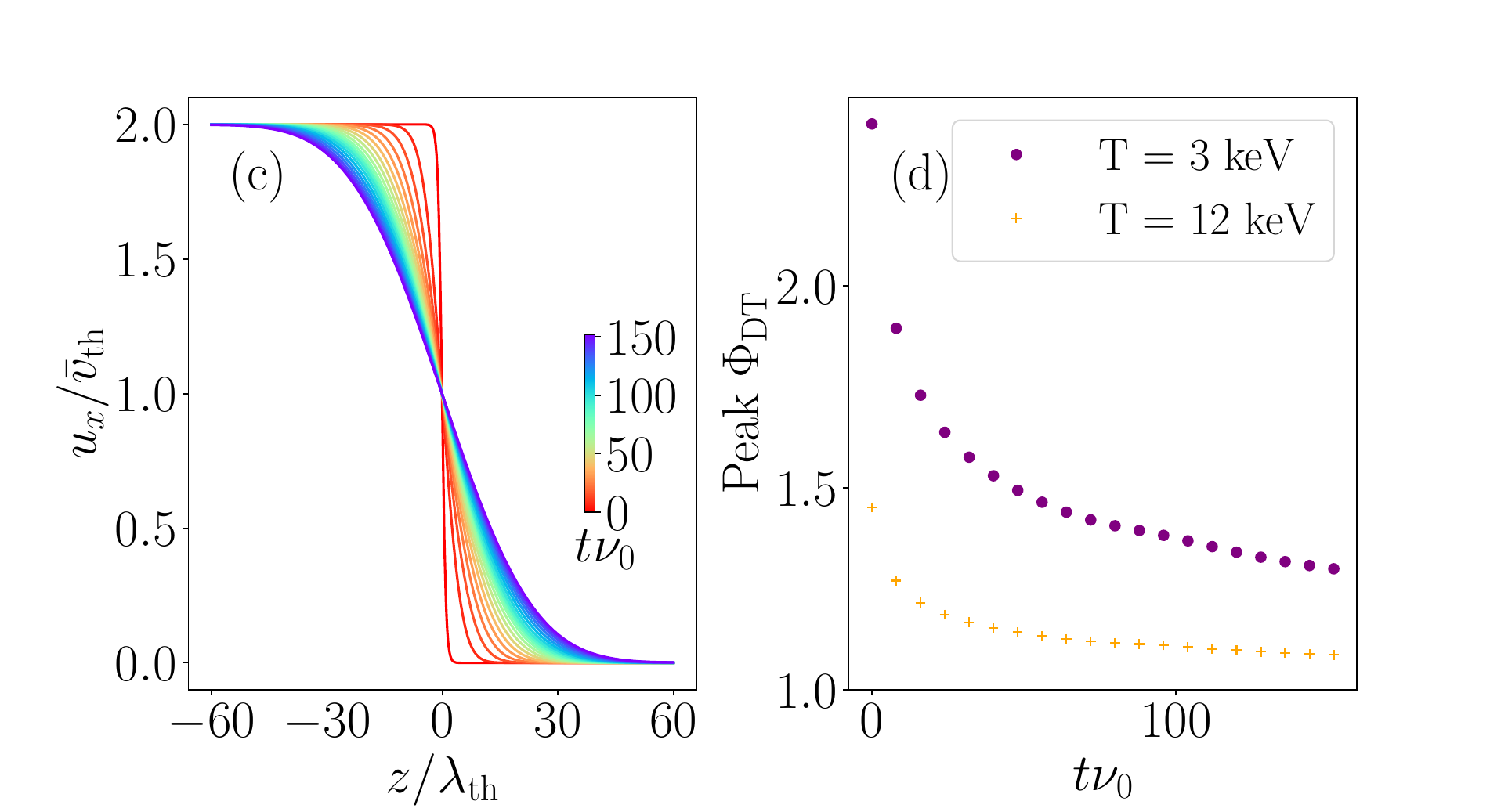}
    \end{subfigure}
    \caption{Top: reactivity enhancement at (a) $T=3~\kev$ and (b) $T=12~\kev$. The left half of each panel shows $\Phi_\mr{DD}$ and the right shows $\Phi_\mr{DT}$. Bottom: (a) viscous relaxation of a shear layer beginning with $u_0=2\vth$ and $L=\lambda_\mr{th}$; and (b) peak $\Phi_\mr{DT}$ at each time step.
    }
    \label{fig_fus_rates_z}
\end{figure}

From these distributions, reactivity was calculated using the cross-section formulas of Bosch \& Hale \cite{Bosch_Hale_1992} for the D(d,n)\textsuperscript{3}He and T(d,n)\textsuperscript{4}He reactions (Fig.~\ref{fig_fus_rates_z}). The enhancements ($\Phi_\mr{DD}$ and $\Phi_\mr{DT}$ respectively) peak just outside the transition layer, but actually dip within the layer because the maximum difference in flow velocity is not as large there as in the outer regions. As anticipated by \eqref{eq_reactivity_approx}, the enhancement factor is larger at low temperature \footnote{Note that the length and velocity scales in each plot depend on temperature. Therefore, the peak is about sixteen times wider, in absolute units, in the $T=12~\kev$ plot.}. 
For one set of conditions, $\Phi_\mr{DT}$ reaches $2.4$ at its peak, a dramatic increase in fusion rate over a region tens of $\lambda_\mr{th}$ in width. 
For equivalent conditions, $\Phi_\mr{DD}$ is larger than $\Phi_\mr{DT}$ due to a leveling off of the DT cross section at high energy. This is consistent with experiments finding a lower than expected DT:DD yield ratio \cite{Johnson_et_2016}. 
The reactivity enhancement is an inherently kinetic effect. 
To illustrate this, moments of $f$ were computed at each $z$ location for one flow profile ($u_0 = \bar{v}_\mr{th}, L = 5\lambda_\mr{th}$) and Maxwellians were generated with the same moments. Reactivities for these Maxwellians are shown in grey in Fig.~\ref{fig_fus_rates_z}, demonstrating that the enhancement, in fact, occurs not only because more energetic ions cross the shear layer, but also because the ions furthest out on the tail are the most affected due to their long mean free paths. 

The flow profile described by \eqref{eq_flow_profile_numerical} is subject to rapid viscous dissipation when $L \sim \lambda_\mr{th}$. To assess the effect of viscosity on reactivity, we performed kinetic simulations in Gkeyll \cite{Hakim_Francisquez_Juno_Hammett_2020} using 
as initial conditions \eqref{eq_flow_profile_numerical} with $u_0 = 2\vth$ and $L = \lambda_\mr{th}$. Fig.~\ref{fig_fus_rates_z} shows the flow profile at later times (c) and the height of the peak of the $\Phi_\mr{DT}$ profile at each time (d). Reactivity drops quickly at first, but begins to level off at later times; ($\Phi_\mr{DT}-1$) drops to about one sixth of its original value in 150 collision times. For the parameters discussed below, this is about $30~\text{ps}$, meaning that the enhancement is significant for a sizable fraction of the burn duration \cite{Kritcher_et_2024}. 
\deletedtext{
This analysis neglects hydrodynamic instabilities, which may transiently increase $\Phi$ even further, much as instabilities aggravate the Knudsen-layer reactivity reduction \cite{Kagan_Svyatskiy_Rinderknecht_Rosenberg_Zylstra_Huang_McDevitt_2015}.
}

\begin{addedblock}
\textit{Shear flow in ICF -- }
As seen in Fig.~\ref{fig_fus_rates_z}d, viscosity is important but does not preclude large effects.  
\end{addedblock}
In magnetic compression \cite{Giuliani_et_2014,Kroupp_et_2011,Maron_2020},
it has been well established that flow kinetic energy can greatly exceed thermal energy; in fact, observations of highly turbulent flows at peak compression in z-pinch experiments \cite{Kroupp_et_2011,Kroupp_et_2018,Davidovits_Kroupp_Stambulchik_Maron_2021} provided inspiration for the consideration of strongly sheared flows in this work.
In laser-driven compression, too, flow kinetic energy can rival or exceed thermal energy \cite{Weber_Clark_Cook_Busby_Robey_2014,Weber_et_2015,Murphy_2014,Clark_et_2015}.
\deletedtext{
Perturbations in the implosion can launch jets of material inward, entering the hot spot with velocities sometimes exceeding $\bar{v}_\mr{th}$ and producing strong shear; 
more disordered turbulent flows are also possible \cite{Grinstein_et_2024,Zhou_Sadler_Hurricane_2025}.
\deletedtext{Quantifying the reactivity enhancement from TKE requires detailed three-dimensional simulations resolving scales below the Gamow mean free path and including a physical viscosity model; these are expensive and hence rare \cite{Haines_et_2020,Grinstein_et_2024}. 
Such simulations have been published for National Ignition Facility (NIF) experiment N120321 \cite{Weber_Clark_Cook_Busby_Robey_2014,Weber_et_2015,Clark_et_2015}. Extracting parameters from these simulations and using the numerical methods described above, we estimate a reactivity enhancement of about $22\%$ in regions near the fill tube jet, averaging to $4\%$ over the entire hot spot. 
Far greater enhancements could be achieved by deliberate driving of small-scale turbulent eddies. }
}\addedtext{A conservative estimate of the size of the effect can be made based on past experiments on the National Ignition Facility (NIF). 
The numerical methods described above, applied to simulations of NIF experiment N120321 \cite{Weber_et_2015,Clark_et_2015}, yield a hot-spot averaged reactivity enhancement of approximately $4\%$. That this is a modest effect is not surprising; NIF targets are carefully designed to minimize turbulence. \addedtext{Nevertheless, even modest changes in reactivity are significant in systems near ignition.} Substantially larger \deletedtext{effects}\addedtext{enhancements} can be contemplated if turbulence is driven deliberately.}
Although viscosity dissipates \addedtext{fine-scale turbulent kinetic energy (TKE)} \deletedtext{TKE} during compression, instabilities \deletedtext{can} provide sudden driving \deletedtext{of flows} at late times;\deletedtext{ as a result} fine-scale TKE has actually been observed to increase near stagnation \cite{Weber_Clark_Cook_Busby_Robey_2014,Zhou_Sadler_Hurricane_2025}. 
In the hot spot at bang time, viscous and inviscid simulations show similar levels of TKE \cite{Weber_Clark_Cook_Busby_Robey_2014,Clark_et_2015}, meaning that viscosity is not playing \deletedtext{a}\addedtext{the} dominant role in quenching turbulence during burn, consistent with Fig.~\ref{fig_fus_rates_z}. It follows that larger TKE could be driven\deletedtext{; how much larger before viscosity becomes insurmountable is an open question. Driving could be done at small ($\mu$m-scale) wavelengths to situate most TKE at reactivity-relevant scales rather than in inert large-scale eddies} \addedtext{at small scales}, perhaps by seeding Rayleigh-Taylor unstable perturbations \cite{Zhou_Sadler_Hurricane_2025} or using structured materials. 
\addedtext{In a $8~\mr{keV}$ plasma, with TKE in micron-scale eddies driven to a level of $90\%$ of the ion thermal energy, $\Phi \approx 2.7$ (this estimate is obtained by the numerical methods above and is detailed in the End Matter). Remarkably, this is comparable to the enhancement if the TKE were thermalized.}


\begin{movedblock}
\ifthenelse{\boolean{showHeadings}}{
    \begin{center}
        \textit{Reduction in ignition energy}
    \end{center}

}{\textit{Reduction in ignition energy -- }}
A corollary \deletedtext{of the effect presented in this Letter}\addedtext{to the shear flow reactivity enhancement effect} is that turbulent plasma may ignite at lower temperature. This leads us to pose the question: \textit{what partition between thermal energy and turbulent kinetic energy (TKE) is optimal for fusion?}
\deletedtext{If reactivity were a purely local, thermal quantity, then the answer would be trivial and any TKE would simply be wasted energy.}\deletedtext{ The fact that the answer is, indeed, nontrivial is of fundamental theoretical interest. }The practical question\deletedtext{, on the other hand,} is whether an ICF design can increase yield by intentionally driving turbulent flows. \addedtext{The following case suggests that huge energy savings are, in fact, possible.}\deletedtext{This question demands intensive further study, but the following case is an extremely promising example.}

\deletedtext{Colder fuel can be ignited}\addedtext{Because ignition can occur in colder fuel} when reactivity is enhanced by TKE\addedtext{, l}\deletedtext{. L}ower electron temperature $T_e$ allows a smaller hot spot because alpha-particle stopping distance scales as $T_e^{3/2}$ \cite{Tabak_et_1994,Tabak_et_2005,Slutz_Vesey_2005}. 
This has remarkable benefits for fast ignition, where a hot spot is heated quickly to initiate fusion in pre-compressed fuel\deletedtext{; with enhanced reactivity, the ignitor can}\addedtext{. If reactivity is augmented by turbulence, then it suffices for the ignitor to} heat a smaller fuel mass to a lower temperature. 
Reducing ignition temperature by a factor of $\tau$ yields a $\tau^{11/2}$ reduction in ignitor energy. 
\addedtext{Now the key result is: \textit{If, for example, enough TKE is driven to permit ignition at $8~\mr{keV}$ rather than at $12~\mr{keV}$, then the required ignitor energy is reduced by a factor of $9.7$. 
Thus, an order-of-magnitude reduction in ignition energy might be achieved in fast ignition.}
}

\deletedtext{If a quiescent hot spot would ignite at $12~\mr{keV}$, the same hot spot can ignite at $10~\mr{keV}$ if TKE is about $30\%$ of thermal energy (see below). 
The energy required to ignite such a hot spot is reduced by $64\%$. Including TKE, the total energy requirement is reduced by $37\%$. 
The ignitor energy is likely more important because TKE can be driven fairly efficiently during compression, whereas fast-ignitor power and intensity requirements present challenges \cite{Tabak_Hinkel_Atzeni_Campbell_Tanaka_2006,Atzeni_2009}.}
\deletedtext{From either perspective, this is a dramatic energy saving. 
Note that this scenario contemplates setting up flows in cold, dense fuel, which is relatively inviscid and so more easily sustains strong shear.}
\addedtext{Moreover, the cold, dense fuel is relatively inviscid until the ignitor pulse heats the hot spot. As ion temperature $T_i$ rises, viscosity increases as $T_i^{5/2}$, but turbulent eddies need only persist for a fraction of the burn time (tens of ps) to facilitate ignition; Fig.~\ref{fig_fus_rates_z}d shows that this is quite reasonable. 
For the example here, assuming a density $\rho = 300~\text{g/cm}^3$, the Gamow mean free path is $\lambda_* \approx 0.8~\mu\mr{m}$; for eddies on this scale, the viscous dissipation time is about 70~ps at $8~\mr{keV}$ but about 12~ns at $1~\mr{keV}$, easily long enough to survive compression.}
\addedtext{Including TKE as well as the ignitor energy, the energy requirement is reduced by a factor of about $6.4$.
This is a simplified model, but it demonstrates from first principles that very large energy savings are possible.}
While localizing TKE and heating in small hot spots comes with its own challenges \cite{Atzeni_1999,Atzeni_2009}, this calculation is broadly a conservative one, ignoring several loss mechanisms that scale favorably with decreasing temperature. 
\deletedtext{Lower ignition temperature comes with other benefits:}\addedtext{For instance,} the final steps of ion heating are the longest and least efficient \cite{Fetsch_Fisch_2023}\addedtext{, while lower $T_e$} \deletedtext{and a lower electron temperature} reduces radiative \deletedtext{and conductive }losses \cite{Atzeni_Meyer-Ter-Vehn_2004,Davidovits_Fisch_2016,Davidovits_Fisch_2019_quasieos}. \movedtext{Many of these favorable scalings apply to conventional ICF schemes as well.}
\end{movedblock}

\begin{movedblock}
\ifthenelse{\boolean{showHeadings}}{
    \begin{center}
        \textit{Discussion}
    \end{center}

}{\textit{Discussion -- }} 
Why has the shear flow reactivity enhancement effect put forth here not already been noted in simulations? The answer lies, at least in part, in its multidimensional nature, involving at least one spatial dimension and two velocity dimensions (see Fig.~\ref{fig_f_panels}), as well as a velocity-dependent collision frequency. Simplifications common in kinetic simulations, such as one-dimensional velocity space, spherical symmetry, and constant collision frequency, obscure these features \cite{Inglebert_Canaud_Larroche_2014,Rinderknecht_Amendt_Wilks_Collins_2018,Hoffman_Zimmerman_Molvig_Rinderknecht_Rosenberg_Albright_Simakov_Sio_Zylstra_GatuJohnson_et,Rosenberg_et_2014,Mannion_et._2023}. 
Nonetheless, this is a fundamental effect relevant to a variety of ICF systems\addedtext{ with likely analogs}\movedtext{ in magnetically confined and astrophysical plasmas.}\deletedtext{.}
\deletedtext{Similar effects are likely relevant}

To date, ICF designs have sought to minimize TKE at bang time on the assumption that turbulence harbors no advantages. This Letter shows that assumption to be false. Remarkably, the sheared flows present in turbulent eddies contribute directly to fusion reactivity. 
\addedtext{The question of optimal partition between turbulent and thermal energy in fusion plasmas is a rich one, offering, beyond its intrinsic theoretical interest, the prospect of novel ICF designs.}
\addedtext{In this Letter, we identified cases where the shear flow reactivity enhancement effect can be large in practical DT implosions.  Of even greater importance, we showed that because ignition then occurs at lower electron temperature, alpha particles are stopped more promptly, leading to a smaller hot spot and vastly decreasing energy requirements.} 
\deletedtext{This raises the question of when, in fact, TKE might be preferable to thermal energy. Fast ignition provides a compelling example, but large savings may be possible elsewhere; for instance, desiging implosions to generate colder, turbulent fuel would reduce radiative losses \cite{Davidovits_Fisch_2016,Davidovits_Fisch_2019_quasieos}. 
We introduced the effect in the context of unmagnetized ICF plasma, but similar effects are likely relevant in magnetically confined and astrophysical plasmas. }

\end{movedblock}

\ifthenelse{\boolean{showHeadings}}{
    \begin{center}
        \textit{Acknowledgments}
    \end{center}

}{\textit{Acknowledgments -- }}
This work was supported by the Center for Magnetic Acceleration, Compression, and Heating (MACH), part of the U.S. DOE-NNSA Stewardship Science Academic Alliances Program under Cooperative Agreement DE-NA0004148.


\bigskip

\newpage
\begin{center}
    \textbf{End Matter}
\end{center}

This End Matter describes in more detail the application of the methods presented in this Letter to calculating the reactivity enhancement in ICF systems. 
For a fill-tube jet entering a hot spot, a simple estimate of the reactivity enhancement can be made by considering a planar shear layer with velocity equal to the jet velocity and a length scale comparable to the size of the shearing region between the jet and background fuel. 
We make such an estimate based on high-resolution viscous simulations \cite{Clark_et_2015,Weber_et_2015} of NIF experiment N120321. 
In a region near the fill-tube jet with ion density $n \sim 10^{25}~\text{cm}^{-3}$ and temperature $T\sim 2.5~\kev$ (cf. Fig.~9 of \cite{Clark_et_2015} and Fig.~7 of \cite{Weber_et_2015}), the simulations indicate $u \sim 1.6\bar{v}_\mr{th}$ and $L \sim 20\lambda_\mr{th}$. DT reactivity is enhanced by $22\%$ in a region about $5~\mu\text{m}$ in width ($\Phi_\mr{DT} \approx 1.22$, while $\Phi_\mr{DD} \approx 1.29$). 
Taking the hot spot radius to be $R\sim 40~\mu\text{m}$ and estimating the total area of the shear layers to be half of the hot spot surface area (although folded deep within the hot spot), the volume-averaged DT reactivity enhancement is a more modest $4\%$. 
In some cases, such as when we are concerned with the initiation of a fusion burn, the local ($22\%$) enhancement is more relevant.

The simple geometry used in the main text to introduce the shear flow reactivity enhancement effect was chosen to highlight the key physics at play. We describe here a simple approximation scheme generalizing the calculations to turbulent flow fields. 
Guided by \eqref{eq_reactivity_approx}, a simple estimate of the reactivity enhancement for a profile of length scale $L$ and normalized flow speed $\hat u$ is
\begin{equation}
    \label{eq_Phi_scaling_CL}
    \Phi \sim 1 + C_L b^{2/3} \hat u^2 ,
\end{equation}
where $C_L$ is an unknown factor depending on $L$. We define a ``Gamow wavenumber''
\begin{equation}
    k_* \doteq \frac{\nu_0}{v_*} \paren{\frac{\vth}{v_*}}^3
\end{equation}
such that sinusoidal flow structures on the scale of $k_*$ have a width comparable to the mean free path of particles at the Gamow peak.

The flow can be broken down into many modes, each with a wavenumber $k$. 
These modes are described by an energy spectrum $E(k)$. Let us assume that the spectral region around $k_*$ can be characterized by a turbulent cascade with power-law form $E(k) \propto k^{-\alpha}$. Although viscosity in ICF hot spots is known to inhibit the development of a true inertial range in many cases \cite{Weber_Clark_Cook_Busby_Robey_2014}, simulations suggest a turbulent cascade at some scales near bang time \cite{Grinstein_et_2024,Zhou_Sadler_Hurricane_2025}. 
All modes contribute to the reactivity, with larger enhancements coming from higher wavenumbers until this trend is arrested by the decreasing energy per mode. We therefore estimate $\Phi$ using the energy in the ${k = k_*}$ mode, \textit{viz.}
\begin{equation}
    \label{eq_Phi_scaling_Cstar}
    \Phi \sim 1 + C_* b^{2/3} \frac{k_* E(k_*)}{(\alpha - 1)E_\mr{th}} ,
\end{equation}
where $E_\mr{th}$ is the \addedtext{ion} thermal energy, $C_*$ is a constant related to $C_L$ and found numerically, and the factor of $k_*/(\alpha - 1)$ estimates the contribution from integrating over all higher-$k$ modes. This underestimates the increase in $\Phi$ at $k > k_*$ as well as the fact that $k < k_*$ modes contribute a nonzero amount to $\Phi$, but integrating over these would require a complicated cutoff procedure at viscous scales; \eqref{eq_Phi_scaling_Cstar} remains a practical, and likely conservative, estimate. Using the method described in the main text, we simulated a shear flow with profile $u(z) = \vth \bs {\hat x} \sin(\pi k_* x)$ in an equimolar DT $3~\mr{keV}$ plasma; taking a spatial average of the resulting DT reactivity yields the estimate $C_* \approx 0.08$ (for DD, $C_* \approx 0.10$). 
The accuracy of \eqref{eq_Phi_scaling_Cstar} was evaluated using simulations of N120321 \cite{Weber_Clark_Cook_Busby_Robey_2014,Weber_et_2015,Clark_et_2015} involving a prominent fill-tube jet. We used kinetic-energy spectra spectra \cite{Weber_et_2015}, which are agnostic of the structure of the jet, along with \eqref{eq_Phi_scaling_Cstar} to compute the reactivity enhancement \addedtext{($u^2 \approx 3\times10^3 \mu \text{m}^2/\text{ns}^2$ at $\lambda_* \approx 1.2 \mu\text{m}$)}. The result of approximately $5\%$ was in good agreement with the estimate based on a planar shear layer.

In many cases, the energy spectrum at small length scales is not known. A simple estimate can be obtained assuming a Kolmogorov spectrum (${\alpha = 5/3}$) and assigning a forcing scale of $k_0$, where most TKE resides. For this model, the enhancement is approximated by 
\begin{equation*}
    \label{eq_Phi_estimate_cascade}
    \Phi \sim 1 + C_* b^{2/3}\frac{E_\mr{TKE}}{E_\mr{th}} \paren{\frac{k_*}{k_0}}^{-2/3} ,
\end{equation*}
where $E_\mr{TKE}$ is the turbulent kinetic energy. This estimate is most reasonable for large $b$, where viscous and Gamow scales are widely separated. It is also reasonable for fast compression where rapid distortion theory applies, and in cases of strong forcing, such that turbulent energy is driven to high wavenumber at rates comparable to viscous damping.

\deletedtext{The End Matter details a scheme for estimating reactivity in such systems.}
\addedtext{To contextualize these formulas, we give two quantitative examples beyond those in the main text.}
\movedtext{At conditions relevant to the onset of burn ($7~\mr{keV}$ and $100~\mr{g/cm}^3$) \cite{Kritcher_et_2024}, with small-scale TKE around $25\%$ of the thermal energy, reactivity is enhanced by about $24\%$ throughout the hot spot.
\movedtext{Here, it is assumed that eddies are on the scale of $2-4\mu\mathrm{m}$; the viscous dissipation time at this scale is $\sim 200~\mr{ps}$.}
\addedtext{The viscous dissipation time $\tau$ is simply estimated by $\tau \sim \widehat L^2/\eta$, where $\widehat L$ is the eddy size normalized to $\lambda_\mr{th}$ and $\eta$ is the kinematic viscosity.}
For a turbulent z-pinch plasma at $4~\mr{g/cm}^3$ and $10~\mr{keV}$, with TKE four times the ion thermal energy and contained in large-scale ($\sim 1\mathrm{mm}$) eddies with a Kolmogorov cascade to smaller scales ($\lambda_* \approx 33~\mu\mathrm{m}$), reactivity is more than doubled ($\Phi \approx 2.3$). }

\bibliography{shear_fusion}

\end{document}